\documentclass[jcp,twocolumn,longbibliography,superscriptaddress]{revtex4-2}
\usepackage{graphicx}
\usepackage[utf8]{inputenc}
\usepackage{amsmath}
\usepackage{verbatim}
\usepackage[labelfont=bf,font=small, justification=raggedright]{caption}
\usepackage{amssymb}
\usepackage{listings}
\usepackage{color}
\usepackage{url}
\usepackage{hyperref}
\usepackage{xspace}
\usepackage{amsmath}
\usepackage{mathtools}
\usepackage{float}
\usepackage{ifthen}
\newcommand{\prog}[1]{Alg.~\ref{alg:#1} (\sub{#1})}

\newcommand{\progg}[1]{Algorithm~\ref{alg:#1} (\sub{#1})}

\newcommand{\progn}[1]{Alg.~\ref{alg:#1}}

\newcommand{\subfig}[2]{{Fig.~\ref{#1}{#2}}} 

\newcommand{\subcap}[1]{{(#1):}}
\newcommand{\pshort}{p^{\text{short}}}
\newcommand{\pishort}{\pi^{\text{short}}}

\newcommand{\pilong}{\pi^{\text{long}}}
\newcommand{\Eq}[1]{Eq.~(\ref{#1})}

\newcommand{\PROCEDURE}[1]{\textbf{procedure}\ \sub{#1}}

\newcommand{\BRACE}[1]{
\;\;\;  \left\{\begin{array}{l}#1\end{array} \right.}
\newcommand{\IS}[2]{#1 \leftarrow #2}

\newcommand{\FOR}[1]{\textbf{for}\ #1 \textbf{: }}
\newcommand{\CHOICE}[1]{\textbf{choice}(#1)}
\newcommand{\WHILE}[1]{\textbf{while } #1 \textbf{: }}
\newcommand{\ENDPROCEDURE}{\text{------} \\ \vspace{-0.8cm}}
\newcommand{\GOTO}[1]{\textbf{goto}\ #1}

\newcommand{\BREAK}{\textbf{break}}

\newcommand{\TRUE}{\textbf{True}}

\newcommand{\IF}[1]{\textbf{if } #1 \textbf{: }}

\newcommand{\ELSE}{\textbf{else: }}

\newcommand{\OUTPUT}[1]{\textbf{output}\ #1}

\newcommand{\INPUT}[1]{\textbf{input}\ #1}
\newcommand{\COMMENT}[1]{\text{\footnotesize (#1)}}

\newcommand{\SET}[1]{\{#1\}}

\newcommand{\sub}[1]{\texttt{#1}}
\newcommand{\Eqtwo}[2]{Eqs~(\ref{#1}) and~(\ref{#2})}
\newcommand{\fig}[1]{Fig.~\ref{#1}}
\newcommand{\figtwo}[2]{Figs~\ref{#1} and \ref{#2}}
\newcommand{\quot}[1]{``#1''}

\newcommand{\app}[1]{Appendix~\ref{#1}}
\newcommand{\sect}[1]{Sec.~\ref{#1}}  

\newcommand{\FCAL}{\mathcal{F}}  
\newcommand{\MCAL}{\mathcal{M}}  
\newcommand{\OCAL}{\mathcal{O}}  
\newcommand{\PCAL}{\mathcal{P}}  
\newcommand{\SCAL}{\mathcal{S}}  
\newcommand{\expa}[1]{\mathrm{e}^{#1}}   
\newcommand{\expb}[1]{\exp \glb #1 \grb} 
\newcommand{\expc}[1]{\exp \glc #1 \grc} 
\newcommand{\ran}{\texttt{ran}}
\newcommand{\ranb}[2][]{\ran_{#1} \! \glb #2 \grb}  

\newcommand{\logb}[2][]{\log^{#1} \glb #2 \grb}  
\newcommand{\logc}[2][]{\log^{#1} \glc #2 \grc}  

\newcommand{\minc}[2][]{\min^{#1} \glc #2 \grc}  

\newcommand{\prob}{\mathbb{P}}

\newcommand{\glb}{\left(}  
\newcommand{\grb}{\right)}  
\newcommand{\glc}{\left[}  
\newcommand{\grc}{\right]}  
\newcommand{\gld}{\left\{}  
\newcommand{\grd}{\right\}}  

\newcommand{\TO}{,\ldots,}
\newcommand{\VEC}[1]{\mathbf{#1}}
\newcommand{\vvec}{\VEC{v}}

\newcommand{\xvec}{\VEC{x}}
\newcommand{\Xvec}{\VEC{X}}

\newcommand{\Yvec}{\VEC{Y}}

\newcommand\bigOb[1]{\ensuremath{\OCAL\glb #1 \grb}}

\newcommand\diff[1]{\mathrm{d}#1}
\newcommand{\fracb}[2]{\frac{#1}{#2}}

\newcommand{\fracd}[2]{\dfrac{#1}{#2}}

\hypersetup{
    colorlinks=true,
    linkcolor=blue,
    filecolor=magenta,      
    urlcolor=cyan,
    pdftitle={Fast long-range sampling},
}

\newcommand{\Ulj}{U^{\text{LJ}}}

\newcommand{\PMet}{\PCAL^{\text{Met}}}
\newcommand{\Pfact}{\PCAL^{\text{fact}}}
\newcommand{\conf}{\Xvec}

\newcommand{\REF}[2][]{
	\ifthenelse{\equal {#1} {}}{Ref.~\cite{#2}}{Ref.~\cite[#1]{#2}}}
\newcommand{\VetoCells}{\SCAL_{\text{veto}}}
\newcommand{\Cell}{\text{Z}}
\newcommand{\FarCells}{\FCAL}
\newcommand{\Atext}{\text{A}}

\newcommand{\Ptext}{\text{P}}
\newcommand{\PPM}{particle-mesh Ewald methods \xspace}
\newcommand{\nshort}{n_s}

\newcommand{\Ushort}{U^{\text{short}}}
\newcommand{\Ulong}{U^{\text{long}}}
\newcommand{\Yold}{\Yvec^{\text{old}}}
\newcommand{\Ynew}{\Yvec^{\text{new}}}

\begin{document}
\newfloat{algorithm}{ht}{loa}
\floatname{algorithm}{Algorithm }
\title{Markov-chain sampling for long-range systems without evaluating
the energy}
\author{Gabriele Tartero}
\email{gabriele.tartero@phys.ens.fr}
\affiliation{Laboratoire de Physique de l’École normale supérieure, ENS,
    Université PSL, CNRS, Sorbonne Université, Université Paris Cité,
Paris, France}
\author{Werner Krauth}
\email{werner.krauth@ens.fr}
\affiliation{Laboratoire de Physique de l’École normale supérieure, ENS,
    Université PSL, CNRS, Sorbonne Université, Université Paris Cité,
Paris, France}
\affiliation{Rudolf Peierls Centre for Theoretical Physics, Clarendon
Laboratory, Oxford OX1 3PU, UK}
\affiliation{Simons Center for Computational Physical Chemistry, New York
University, New York (NY), USA}

\date{\today}

\begin{abstract}
In past decades, enormous effort has been expended to develop algorithms and
even to construct special-purpose computers in order to efficiently evaluate
total energies and forces for long-range-interacting particle systems, with
the particle-mesh Ewald and the fast multipole methods as well as the
\quot{Anton}
series of supercomputers serving as examples for biomolecular simulations.
Cutoffs
in the range of the interaction have also been used for large systems. All these
methods require extrapolations.
Within Markov-chain Monte Carlo, in thermal equilibrium, the Boltzmann
distribution can however be sampled natively without evaluating the total
interaction potential. Using as an example the Lennard-Jones interaction,
we review past attempts in this direction, and then
discuss in detail the class of cell-veto algorithms which make possible
fast, native sampling of the Boltzmann distribution without any approximation,
extrapolation, or cutoff even for the slowly decaying Coulomb interaction. The
computing effort
per move remains constant with increasing system size, as we show explicitly.
We provide worked-out illustrations and pseudocode representations of the
discussed algorithms. Python implementations are made available in an
associated open-source software repository.
\end{abstract}

\maketitle

\section{Introduction}

One of the aims of computational science is to analyze, and often solve, model
systems in fields from subatomic particles to condensed-matter physics and
chemistry, to galaxies and to the universe. The power of modern computers
appears without limits when compared to the early electronic devices from only
two generations ago. Algorithms have also much developed, as has the ease with
which computers are interfaced and their output processed. Nevertheless, certain
limits persist. The first, broadly speaking, is a limitation in space: One tries
to simulate large, possibly infinite systems, but is restricted by the finite
extent of computer memories. Second, one may aim for long-time simulations but
the necessary evolution equations may not allow one to reach the relevant
temporal scales. The third limitation is in the type of couplings between
particles or fields. They may be of many-body nature, including
quantum and long-range potentials and may require even today
require simplifications~\cite{Schlick2002}.

In classical condensed-matter particle systems in thermal equilibrium, the
limitations on space, time, and what could be called \quot{range} can again be
illustrated. In contexts from statistical mechanics to  molecular simulation in
biochemistry, one may want to analyze large, almost infinite systems, but only
in exceptional cases is it possible to simulate them directly
~\cite{EissfellerOpper1992,Jordan2008,VanHoucke1,VanHouckeEOS,RossiCDet}.
Prominent coarse-graining strategies were developed~\cite{Levitt2014} in order
to reach larger and larger sizes. On the other hand, it is the change of
behavior of finite systems with size, the famous finite-size scaling, which
often provides crucial information on a physical model~\cite{Cardy1996}. One
also may want to reach essentially infinite simulation times, as the
equilibrium state is approached from an initial configuration only in this
limit. Computer computations are by definition of finite duration and, within
Markov-chain Monte Carlo, it is only in exceptional cases possible to reach the
infinite-time limit directly~\cite{AldousDiaconis1986,ProppWilson1996}.

The approximation of complex interaction potentials, the above \quot{range}
limitation, is what we are concerned with in this paper, again in the framework
of equilibrium physics. We consider systems of particles at a given temperature,
and governed by the Boltzmann distribution. Such systems can be simulated by
molecular dynamics or sampled by the Monte Carlo method. Particle systems with
Coulomb or Lennard-Jones interactions will serve as examples. The former decay
as $1/r$ with the distance $r$ between charged particles. Its long-range nature
derives from the masslessness of the photon. The Lennard-Jones potential
describes the interaction between uncharged, non-polar atom. Repulsive at small
distances, its attractive $1/r^6$ behavior at large distances describes the
London dispersive force~\cite{Stone2013}. For both the Coulomb and the
Lennnard-Jones potentials, the long-range nature is thus rooted in profound
physical principles, and there is a strong incentive to maintain it in the
modeling. However, long-range potentials pose severe problems for computation,
which arise, naively, because the total energy of a system of $N$ particles
interacting in pairs consists of $N(N-1)/2$ terms. Moving one particle changes
$N-1$ terms in the total energy. Likewise, the force on a given particle is
composed of $ N -1$ terms.

Computational science, in the past decades, has focused on how to
evaluate long-ranged potentials or forces. For the case of the Coulomb
interaction, the potential or the forces are computed by interpolating charges
to a grid, then solving Poisson's equation in discretized momentum space using
fast Fourier transform. The \quot{Anton} series of supercomputers has been
designed in order to optimize the interpolation and force
evaluation~\cite{ShawAnton2007}, yielding spectacular
speedups~\cite{Shaw2010}. The discretization error of these so-called \PPM
disappears only for vanishing grid size. The fast multipole
method~\cite{greengard1987} presents another extrapolation of the interaction in
terms of moments of the charge distribution. Very often also, the Coulomb
interaction is cut off beyond a certain radius, although this profoundly
modifies the underlying physics of the models. Discretization and cutoffs both
call for extrapolations. For decades, the Lennard-Jones potential was cut off at
a finite distance, although artifacts introduced by the cutoff on phase
boundaries~\cite{Smit1991,Smit1992} and on interface
effects~\cite{LeiBykov2005, Troester2012} were pointed out repeatedly. In
recent years,
this problem has been identified, and the particle-mesh Ewald approach extended
to the Lennard-Jones system~\cite{Tempra2022}. However, this creates an
avalanche of problems, as the empirical interaction potentials used for
the sampling are themselves fitted (originally with a cutoff) and then have to
be reparametrized~\cite{Yu2021}.

A Monte Carlo algorithm typically consists in a sequence of simple proposed
moves, that are sampled from a certain probability distribution, and that are
either accepted or rejected~\cite{SMAC}. There is much liberty in which moves
to propose, and how to accept or reject them with a \quot{filter}. The only
condition to be satisfied is that the Boltzmann distribution be stationary with
respect to the move set encoded in a transition matrix. As the proposed move is
simple, Markov-chain Monte Carlo is a decision problem in nature.

In this paper, we discuss methods for sampling the Boltzmann distribution
$\expb{-\beta U}$ without evaluating $U$. This approach is conceptually distinct
from  the previously mentioned computational focus. It leads to the fast
\emph{native} sampling (without any cutoffs, discretizations or time-stepping
errors) for long-range potentials. In the main part of the paper, we discuss in
detail the cell-veto class of algorithms\cite{KapferKrauth2016}, which rely on
two key ingredients: First, the aforementioned accept/reject decision is
split, using the factorized Metropolis filter~\cite{Michel2014JCP}, into a
large number of independent factor decisions, which are coordinated through a
consensus principle rather than through the computation of the system energy or
its derivatives. Second, the accept/reject decision for a factor is generally
undertaken in two steps using two \quot{pebbles}~\cite{Tartero2024} that
correspond to the thinning of an inhomogeneous Poisson
process~\cite{LewisShedler1979}: first, in an approximate (conservative)
fashion, that provisionally rejects \quot{too many moves}, and, second, in a
correction
step that accepts some of the provisional rejects such that the overall
rejection probability is correct.  We provide a step-by-step introduction to
this class of algorithms, together with heuristics, pseudocodes and simplified
Python implementations. We thus hope to facilitate access to a class of
algorithms which have remained cryptic and have never been presented in the
context of decision problems and their possible
extensions~\cite{MuellerJanke2023,Berne2002,
MichelTanDeng2019,Hoellmer2023molecular}.

\section{From long-range potentials to the cell-veto algorithm}

We concentrate on long-range pair potentials, leaving aside many-body
terms that can be treated as well~\cite{Harland2017,Hoellmer2023molecular}.
Several strategies for treating the long-range nature are discussed in
\sect{subsec:physical_model}, such as solving the accept/reject decision with
approximate potentials~\cite{MuellerJanke2023}, and as factorizing the total
energy into short-range and long-range contributions~\cite{Berne2002}. The
factorized Metropolis filter generalizes this approach
(\sect{subsec:consensus_principle}). Together with the concept of bounding
potential,  it lies at the heart of the native \bigOb{1} cell-veto algorithms,
of which we present a naive version in \sect{subsec:bounding_potential} before
sharpening it in \sect{sec:constant_complexity}.

\subsection{Potentials, Metropolis algorithm}
\label{subsec:physical_model}

\begin{figure}[htb]
	\centering
	\includegraphics{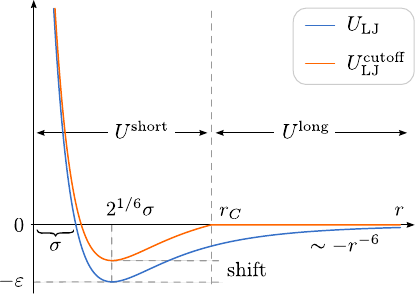}
	\caption{Lennard-Jones potential with short-range and long-range parts
separated by a cutoff $r_c$ and a shifted cutoff variant which vanishes beyond
$r_c$. For large $r$, $\Ulj(r) \sim r^{-6}$ is attractive.}
	\label{fig:LennardJonesPotential}
\end{figure}

Specifically, we consider throughout this paper $N$ particles
inside a two-dimensional periodic box of size $L$ interacting \emph{via} the
Lennard-Jones interaction~\cite{LennardJones1931}:
\begin{equation}
	\Ulj(r) = 4 \varepsilon \glc \glb \fracb{\sigma}{r} \grb ^{12} - \glb 
	\fracb{\sigma}{r} \grb ^{6} \grc, 
	\label{equ:lj_potential}
\end{equation}
At large distances, its attractive tail vanishes as $1/r^6$ (see
\fig{fig:LennardJonesPotential}). 
Our sample programs are for this
two-dimensional system, but they generalize easily to higher dimensions and
to arbitrary long-range interactions (see \app{app:Software}).
For the most part, a cutoff means that one can divide the potential $U$
into a short-range and a long-range contribution:
\begin{equation}
 U(r) = \underbrace{ U(r) \Theta(r_c - r)}_{\Ushort} +
\underbrace{ U(r) \Theta(r - r_c)}_{\Ulong},
\label{equ:ShortLongDecomposition}
\end{equation}
where $\Theta(z)$ is the unit-ramp function, which is zero for negative $z$ and
one for positive $z$. 
The Boltzmann distribution of the potential $U$ then
becomes the product of Boltzmann distributions of the constituents:
\begin{equation}
\pi(\Xvec) = \pishort(\Xvec) \pilong(\Xvec).
\label{equ:ShortLongFactors}
\end{equation}

Commonly, a cutoff replaces $U(r)$ by its
short-range part, and the long-range part is set to zero (see
\fig{fig:LennardJonesPotential}, for the Lennard-Jones potential). In that case,
it is customary to shift $\Ushort$, so that the approximate potential is
continuous. More elaborate procedures are common in order to render the
approximate potential better behaved around $r_c$, as often required for
molecular dynamics.

\begin{figure}[htb]
    \centering
    \includegraphics{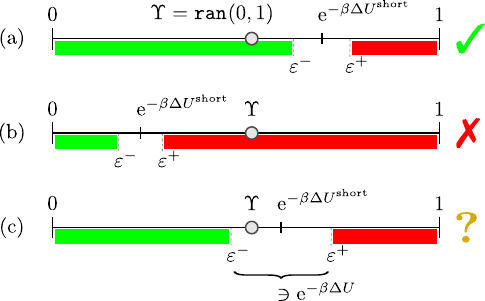}
    \caption{Decisions in MCMC. In the Metropolis filter of
    \Eq{equ:metropolis_filter}, the accept/reject decision for
$\expa{-\beta U}$ can often be made from an approximation $\expa{-\beta
\Ushort}$ with
error bounds. }
    \label{fig:MetropolisDecision}
\end{figure}

\begin{figure}[htb]
    \centering
    \includegraphics{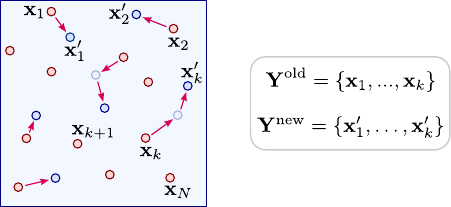}
\caption{Multi-step Metropolis algorithm. In \progn{multi-step-metropolis},
moves are first made with $\Ushort$ only. Initial and final positions are
stored. The proposed move $\Yold \to \Ynew$ is subject to the Metropolis filter
with $\Ulong$.}
\label{fig:BerneSchema}
\end{figure}

In the Metropolis algorithm, a random particle at position $\xvec \in \conf$ is
often selected for random displacement to  $\xvec' = \xvec + \Delta \xvec$.
The move from $\Xvec $ to $\Xvec'$ (in the latter, $\xvec$ is replaced by
$\xvec'$) is
accepted with a probability given by the Metropolis filter
\begin{equation}
    \PMet(\Xvec \to \Xvec')
     = \min \glc 1 , \expb{-\beta \Delta U} \grc.
\label{equ:metropolis_filter}
\end{equation}
With a system-size independent cutoff, the number of factors that have to be
considered for each move remains constant. These factors can be identified
with a standard cell system  and evaluated in \bigOb{1} operations. However, the
essential long-range nature of the potential is then sacrificed.

Strategies have been devised to improve the \bigOb{N} scaling of
the Metropolis algorithm natively. One natural approach exploits the
decision-problem nature of the Metropolis filter. The Bernoulli distribution
giving rise to \Eq{equ:metropolis_filter} is sampled with a uniform random
number $\ranb{0,1} = \Upsilon$, and the move is accepted if $\Upsilon <
\expb{-\beta U}$ and rejected otherwise. For a given dynamical cutoff $r_c$, the
accept/reject decision may be taken on the basis of $\Ushort(r_c)$
if the neglected long-range contribution  $\Ulong(r_c)$ can be proven not to
change the decision (see \fig{fig:MetropolisDecision}). The cutoff $r_c$ is
then increased if required, adding particles in concentric layers around $\xvec
$ or $\xvec'$. Under mild conditions on the maximum local density, this approach
can be made
rigorous~\cite{MuellerJanke2023}. However, the sheer number of evaluations
(which can
easily reach $10^{10}$ or $10^{12}$) renders it difficult to implement.
Also, dynamical cutoffs do not allow for long-range potentials which, in
$d$-dimensional space, decay as $~r^{-d}$ or slower, as the contribution of
$\Ulong$ then remains unbounded for large
$r_c$~\cite{KapferKrauth2016,Faulkner2018}.

In the multi-time-step Metropolis algorithm~\cite{Berne2002}, the Lennard-Jones
system again splits the potential $U$ into long-range and short-range
contributions, performing  a series of $n_s$ provisional moves on the basis
of  $\Ushort$ only. This amounts to proposing a move of $k<N$
particles (with coordinates $\Yvec$, see \fig{fig:BerneSchema}) from
$\Yvec^{\text{old}}$ to $\Yvec^{\text{new}}$.
These \quot{short-range} moves are then accepted/rejected with the Metropolis
filter for $\Ulong$ only.
Calling for simplicity the original configuration
$\Xvec$ (with the $k$ particles at $\Yold$) and the target configuration
$\Xvec'$ (with the $k$ particles at $\Ynew$), the composite probabilities are
\begin{equation}
\begin{aligned}
p(\Xvec \to \Xvec') &= \pshort (\Xvec \to \Xvec') \minc{1,
\fracb {\pilong(\Xvec')} {\pilong(\Xvec)}}
\\
p(\Xvec' \to \Xvec) &= \pshort (\Xvec' \to \Xvec) \minc{1,
\fracb {\pilong(\Xvec)} {\pilong(\Xvec')}}.
\end{aligned}
\label{equ:BerneDetailed1}
\end{equation}
Taking the ratio in \Eq{equ:BerneDetailed1} and noting
that $\pshort$, as a sequence of $n_s$ reversible moves, satisfies
the detailed-balance condition with respect to $\pishort$:
\begin{equation*}
\pishort(\Xvec) \pshort(\Xvec \to \Xvec') =
\pishort(\Xvec') \pshort(\Xvec' \to \Xvec),
\end{equation*}
we see that the full transition matrix satisfies detailed balance with respect
to the full Boltzmann distribution $\pi = \pishort \pilong$
\begin{equation}
\pi(\Xvec)  p(\Xvec \to \Xvec')  = \pi(\Xvec') p(\Xvec' \to \Xvec),
\label{equ:BerneDetailedFull}
\end{equation}
so that the algorithm is correct (see  \prog{multi-step-metropolis} for an
implementation).

In \prog{multi-step-metropolis}), short-range moves can be implemented in
\bigOb{1} with an appropriate  cell system~\cite[Sect. 2.4.1]{SMAC}). The
construction of the sets $\Yold$ and $\Ynew$ is also \bigOb{1} per element, so
that the complexity of the entire short-range loop is \bigOb{n_s}. However, the
complexity of the long-range decision in \progn{multi-step-metropolis} is
\bigOb{n_s \times N}, with an acceptance probability that plummets with
increasing $n_s$, a parameter that should therefore be chosen to be as small as
possible. This leads to $n_s = 1$, and this defeats the initial intention.

\begin{algorithm}
    \newcommand{\algo}{multi-step-metropolis}
    \captionsetup{margin=0pt,justification=raggedright}
    \begin{center}
        $\begin{array}{ll}
            & \PROCEDURE{\algo}\\
            & \INPUT{\conf}\ \COMMENT{configuration at time
$t$}\\
            & \IS{\Yold}{\emptyset};\ \IS{\Ynew}{\emptyset}\ \\
            & \FOR{i = 1 \TO \nshort}\ \COMMENT{short-range steps}\\
            & \BRACE{
                     \IS{\xvec}{\CHOICE{\Xvec}}\ \COMMENT{random
particle}\\
                     \IS{U^{\text{short}}_\xvec}{\sum_{\xvec'' \in \Xvec
\setminus \SET{\xvec}: | \xvec'' - \xvec| < r_c} U(|\xvec''- \xvec|)} \\
                     \IS{\xvec'}{\xvec + \Delta \xvec}\ \COMMENT{with $|\Delta 
                     \xvec| < \delta$} \\
                     \IS{U^{\text{short}}_{\xvec'}}{\sum_{\xvec'' \in \Xvec
\setminus \SET{\xvec}: | \xvec'' -
\xvec'| < r_c} U(|\xvec''- \xvec'|)} \\
                     \IS{\Upsilon}{\ranb{0, 1}}\\
                     \IF{\Upsilon < \expc{-\beta \glb U_{\xvec'}^{\text{short}}
- U_\xvec^{\text{short}} \grb }} \\
                    \BRACE{
  \IS{\conf}{\SET{\xvec'} \cup \conf \setminus \SET{\xvec}}; \IS{\Ynew}{\Ynew
\cup \SET{\xvec'}}\\
  \IF{\xvec \not \in \Ynew} \IS{\Yold}{\Yold \cup
\SET{\xvec}} \\
   \ELSE \IS{\Ynew}{\Ynew \setminus \SET{ \xvec}} \\
   }\\
}\\
            & \IS{\Delta U^{\text{long}}}{\!\!\!\!\!\!\!\sideset{}{'}\sum
\limits_{\stackrel{\!\!\!\xvec \in \Ynew, \xvec' \in \Xvec \!\!\!}{| \xvec -
\xvec'| > r_c}} \!\!\!\!\!\!\! U(|\xvec - \xvec'|)
 -
 \!\!\!\!\! \!\!\!\!\!
\!\!\!\! \sideset{}{'}\sum\limits_{\stackrel{\!\!\!\xvec \in \Yold, \xvec'
\in \Yold \cup \Xvec
\setminus \Ynew\!\!\!}{| \xvec - \xvec'| > r_c}} \!\!\!\! \!\!\! \!\!
U(|\xvec - \xvec'|)} \\
            & \IS{\Upsilon}{\ranb{0, 1}}\ \COMMENT{long-range decision}\\
            & \IF{\Upsilon > \expb{-\beta \Delta U^{\text{long}}}} \\
            & \BRACE{ \IS{\Xvec}{\Yold \cup \Xvec \setminus \Ynew}
            }\\
            & \OUTPUT{\conf}\ \COMMENT{configuration at time $t+1$}\\
            & \ENDPROCEDURE\
        \end{array}$
    \end{center}
    \caption{\sub{\algo}. Composite iteration of the algorithm of
    \REF{Berne2002}. The calculation of $U_\xvec^{\text{short}}$ is
   \bigOb{1} with the use of a grid. The prime in $\sum'$ eliminates double
counting    of pairs $(\xvec, \xvec')$ and $(\xvec', \xvec)$.
For large $n_s$, this algorithm has many long-range rejections.}
    \label{alg:\algo}
\end{algorithm} 

\progg{multi-step-metropolis}, in spite of its serious limitations, illustrates
the split of $U$ into $\Ushort(r)$ and $\Ulong(r)$ that effectively leads to a
factorization of the interaction. The concept will be extended in the following
sections, where it leads to native \bigOb{1} long-range algorithms. The split of
the interactions also echoes Hamiltonian Monte
Carlo~\cite{DuKePeRo1987,Neal2011,BouRabee2018},
where provisional moves $\Xvec \to \Xvec'$ are proposed from a
sequence of molecular-dynamics iterations rather than from a number of
Metropolis steps. In Hamiltonian Monte Carlo, accumulated finite-time-step
errors are also eliminated by a Metropolis filter, and the Boltzmann
distribution is again sampled without approximations. This works because the
errors are small, but it fails for large long-range systems because the errors
are extensive in $N$ even for $n_s=1$.

\subsection{Factorized Metropolis filter}
\label{subsec:consensus_principle}

The multi-time-step Metropolis appoach of \progn{multi-step-metropolis}
separates
the interactions into two factors, one short-range and one long-range.
More generally, potentials $U$ can often be written as
\begin{equation}
    U(\conf) = \sum_{M \in \MCAL} U_M,
\label{equ:FactorizationPotential}
\end{equation}
with factors $M$ in a set $\MCAL$, so that the Boltzmann distribution is
\begin{equation}
    \pi(\conf) = \expc{-\beta U(\conf)} = \prod_{M \in \MCAL} \expb{-\beta
    U_M}.
\label{equ:boltzmann_distribution}
\end{equation}
We can then replace the Metropolis filter of \Eq{equ:metropolis_filter}, written
as
\begin{equation}
\PMet(\xvec \to \xvec')
     = \min \glc 1 , \prod_{M \in \MCAL}\expb{-\beta \Delta U_M} \grc,
\label{equ:metropolis_filterFactor}
\end{equation}
with the factorized Metropolis filter~\cite{Michel2014JCP}
\begin{equation}
     \Pfact(\xvec \to \xvec')  = \prod_{M \in \MCAL}
     \underbrace{\min \glc 1 , \expb{- \beta \Delta U_{M}}
     \grc}_{\Pfact_{M}(\xvec \to \xvec')}.
     \label{equ:factorized_filter}
\end{equation}
That the factorized filter also satisfies the detailed-balance condition
can be shown just as in \Eqtwo{equ:BerneDetailed1}{equ:BerneDetailedFull}.

\begin{figure*}[tb]
    \centering
    \includegraphics{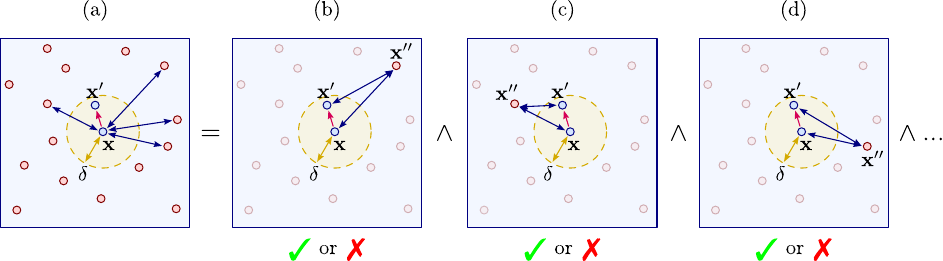}
    \caption{Factorized Metropolis filter using the consensus
    principle (see \progn{factorized-metropolis}).
    \subcap{a} The active particle $\xvec$ may interact with all other
    particles.
    \subcap{b,c,d, \dots} Every factor of the factorized Metropolis filter in
    \Eq{equ:factorized_filter} independently decides whether to veto
    a proposed move. The conjunctions $\wedge$ denote the consensus:
    the move $\xvec \to \xvec'$ is accepted if no factor vetoes it.
    }
\label{fig:FactorizedMetropolis}
\end{figure*}

In our context, the use of pair factors $M = \SET{i,j}$ with $U_M = U(|\xvec_i -
\xvec_j|)$ is natural (even though the concept is easily
generalizable~\cite{Michel2014JCP,Faulkner2018,Tartero2024}). A proposed move of
a single particle then involves $N-1$ factors (that is, $N-1$ pairs). It is
accepted by consensus, that is, if all factors accept it (see
\prog{factorized-metropolis} and \fig{fig:FactorizedMetropolis}; see
\REF{Tartero2024} for a general discussion of the consensus principle). We will
show below how to implement this by evaluating \bigOb{1} pairs, thus with
only \bigOb{1} operations.

\begin{algorithm}
    \newcommand{\algo}{factorized-metropolis}
    \captionsetup{margin=0pt,justification=raggedright}
    \begin{center}
        $\begin{array}{ll}
            & \PROCEDURE{\algo}\\
            & \INPUT{\conf}\ \COMMENT{configuration at time $t$}\\
            & \IS{\xvec}{\CHOICE{\conf}}\ \COMMENT{random particle}\\
            & \IS{\xvec'}{\xvec +
            \Delta \xvec}\ \COMMENT{with $|\Delta \xvec| < \delta$} \\
            & \FOR{\xvec''  \in \conf \setminus \SET{\xvec} }\\
            & \BRACE{\IS{\Upsilon}{\ranb{0, 1}}\\
                     \IF{\Upsilon > \expc{-\beta \glb
                     U_{\xvec'' \xvec'}  - U_{\xvec'' \xvec}
\grb  }} \GOTO 1
                    }\\
             & \IS{\conf}{\SET{\xvec'} \cup \conf \setminus \SET{\xvec}} \\
          1 & \OUTPUT{\conf}\ \COMMENT{configuration at time $t+1$}\\
            & \ENDPROCEDURE\
        \end{array}$
    \end{center}
    \caption{\sub{\algo}. Proposing a move, and accepting it in case of
   consensus, also samples the Bolzmann distribution of
\Eq{equ:boltzmann_distribution}.}
\label{alg:\algo}
\end{algorithm}

\subsection{Bounding the potential, cell-veto algorithm}
\label{subsec:bounding_potential}
The factorized Metropolis filter replaces the accept/reject decision based on
the total energy $U(x)$ with independent decisions for all factor
potentials $U_M$ with $M \in \MCAL$. This
replacement
carries enormous potential for speedup as, for a given distant pair of
particles $\xvec$ and $\xvec''$, we need not systematically evaluate all the
rejection probabilities~\cite{KapferKrauth2016}, applying a principle called
\quot{thinning}~\cite{LewisShedler1979}. Heuristically, let us suppose that, at
a distance between $\xvec $ and $\xvec ''$, the veto probability $q =
1-\expb{-\beta \Delta U_M}$ can be bounded by $\epsilon \ll 1$. Rather than
check the factor at each proposed move, we check it with probability
$\epsilon$ (on average every $1/\epsilon$ times). Only when we check it, we
evaluate $q$, and veto with probability $q / \epsilon$. Because of $q =
\epsilon \times (q / \epsilon)$, the overall veto probability is correct, but
the bound has allowed us to drastically reduce the number of evaluations.

\begin{figure}[htb]
    \centering
    \includegraphics{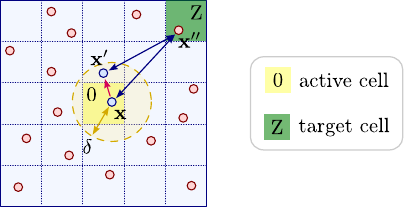}
    \caption{For a pair of far-away particles $\xvec$ and $\xvec''$, the
factor rejection probability for a move $\xvec \to \xvec'$
can be bounded by a constant $q_\Cell$ depending on the target
cell $\Cell$ relative to the active cell $0$.}
    \label{fig:BoxCells}
\end{figure}

The cell-veto algorithm~\cite{KapferKrauth2016} puts the above heuristic into
practice by dividing the periodic simulation box into cells that usually contain
at most one particle (see \fig{fig:BoxCells}). Rather than the particle
$\xvec''$, one addresses a target cell $\Cell$ relative to the active cell $0$
containing the active particle $\xvec$. The upper bound $q_\Cell$ for the
rejection plays the role of $\epsilon$ in the heuristic. It satisfies:
\begin{equation}
q_\Cell \ge \!\!\!\! \max_{
\substack{
\xvec \in 0, \xvec'' \in \Cell
\\
\xvec': |\xvec'- \xvec| < \delta
}
}
\!\!\!
\gld
 1 -
\expc{-\beta \glb U_{\xvec'' \xvec'}  - U_{\xvec'' \xvec}
\grb } \grd .
\end{equation}
`To yield the correct rejection probability
$\glc 1 - \expc{-\beta \glb U_{\xvec'' \xvec'}  - U_{\xvec'' \xvec}
\grb } \grc
{q_\Cell}$,
a veto called  by cell $\Cell$, with probability $q_\Cell$, must be confirmed
with probability
\begin{equation}
\fracb{
\glc
 1 - \expc{-\beta \glb U_{\xvec'' \xvec'}  - U_{\xvec'' \xvec}
\grb } \grc}
{q_\Cell},
\end{equation}

\begin{figure*}[htb]
    \centering
    \includegraphics{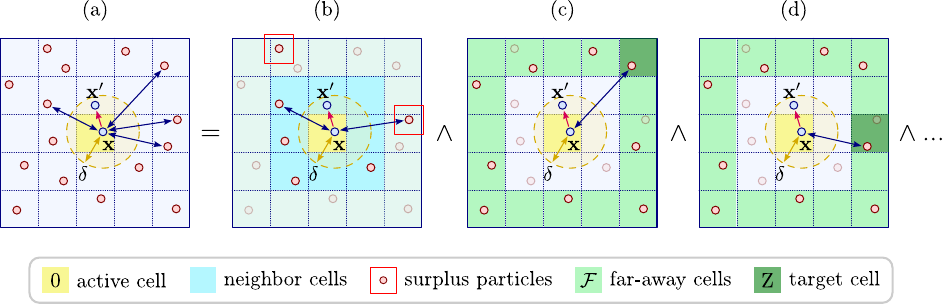}
    \caption{Naive cell-veto algorithm for a move $\xvec \to \xvec'$, as
implemented in \prog{cell-veto}. \subcap{a} The box is divided into cells which
rarely contain more than one particle. \subcap{b} Neighbor and surplus particles
are handled directly, as in \progn{factorized-metropolis}. \subcap{c,d, \dots}
Cell-veto loop. The far-away cells $\FarCells$ are iterated over. Each cell
$\Cell$, then the particle $\xvec''$ inside it, may veto the move.
}
    \label{fig:CellVeto}
\end{figure*}

Two complications must be solved. First, a cell $\Cell$ neighboring the active
cell $0$  does not allow for a finite upper bound $q_\Cell$. Second, the
cell $\Cell$ may contain more than one particle. Both cases can be easily
treated
with the second requiring a set of \quot{surplus} particles. The
\quot{two-pebble} decision~\cite{Tartero2024}, first for the cell $\Cell$, then
for the particle $\xvec'' \in \Cell$, is illustrated in \prog{cell-veto}, which
also implements a first loop over neighbor and surplus particles, as in
\prog{factorized-metropolis}, and a cell-veto loop over far-away cells that form
a set $\FarCells$ (see \fig{fig:CellVeto}). This loop over the elements of
$\FarCells$ is naive.

\begin{algorithm}
    \newcommand{\algo}{cell-veto}
    \captionsetup{margin=0pt,justification=raggedright}
    \begin{center}
        $\begin{array}{ll}
            & \PROCEDURE{\algo}\\
            & \INPUT{\conf}\ \COMMENT{configuration of  $N$ particles}\\
            & \IS{\xvec}{\CHOICE{\conf}} \\
            & \IS{\xvec'}{\xvec + \Delta\xvec}\ \COMMENT{with $|\Delta \xvec| <
\delta$}\\
            & \FOR{\xvec'' \in \SET{\text{neighbor}} \cup
\SET{\text{surplus} }}\ \COMMENT{see 
\subfig{fig:CellVeto}{b}}\\
            & \BRACE{\IS{\Upsilon}{\ranb{0, 1}}\\
 \IF{\Upsilon > \expc{-\beta \glb U_{\xvec'' \xvec'} - U_{\xvec'' \xvec}
\grb }} \GOTO 1 \\
               }\\
            & \FOR{\Cell \in \FarCells, \text{non-empty}}\
             \COMMENT{loop over far-away cells $\FarCells$} \\
            & \BRACE{
            \IS{\Upsilon_1}{\ranb{0, 1}}\\
                    \IF{\Upsilon_1 > q_\Cell }\\\
                    \BRACE{
                    \IS{\Upsilon_2}{\ranb{0, 1}}\\
                    \IS{\xvec''}{\text{particle in cell $\Cell$}} \\
                    \IF{\Upsilon_2 < \fracd{1 -
\expc{-\beta \glb U_{\xvec'' \xvec'}  - U_{\xvec'' \xvec}
\grb }
                                    }{q_\Cell
}}  \GOTO 1 \\
                    }\\ 
                    }\\ 
            &\IS{\conf}{\SET{ \xvec'} \cup \conf \setminus \SET{\xvec}} \\
           1 & \OUTPUT{\conf}\\
            & \ENDPROCEDURE\
        \end{array}$
    \end{center}
    \caption{\sub{\algo}. Naive cell-veto algorithm. The far-away cells $\Cell$
    use a two-pebble veto, one for the cell, one for the particle
(see patch in \progn{cell-veto(patch)}).}
\label{alg:\algo}
\end{algorithm}

\section{Cell-veto loop}
\label{sec:constant_complexity}

The loop over the far-away cells $\FarCells$  in the naive \prog{cell-veto}
scales linearly with $N$. It can be avoided by sampling the subset
$\VetoCells$ of cells vetoing the proposed move
(\sect{subsec:powerset_sampling}). Two different scenarios are possible. For
infinitesimal moves, when the cell-veto algorithm is implemented for a
continuous-time Morkov process, as in the event-chain Monte Carlo
algorithm~\cite{Krauth2021eventchain}, the veto probabilities are themselves
infinitesimal and, most of the time, the set $\VetoCells$ is empty. Otherwise,
if $\VetoCells \neq \emptyset$,
this set contains a single cell, which can be sampled in \bigOb{1} using
Walker's
algorithm~\cite{Walker1977AnEfficientMethod} (\sect{subsec:event-chain}). In
contrast, for finite moves, the set $\VetoCells$ may contain more than one
element, in which case it can be sampled  without iterating over all
cells~\cite{MichelTanDeng2019}. We present an algorithm which reduces the
sampling problem to the infinitesimal-move case
(\sect{subsec:powerset-poisson}).

\subsection{The set $\VetoCells$ of veto cells}
\label{subsec:powerset_sampling}
In \prog{cell-veto}, all cells are looped over, but the veto must be confirmed
in a second step only for the subset $\VetoCells \subset \FarCells$ of vetoing
cells. The \quot{cell-accept} decisions are confirmed
automatically~\cite{Tartero2024}. Clearly, the probability of any such subset is
\begin{equation}
    \pi(\VetoCells) = 
    \underbrace{\prod_{\Cell \in \VetoCells}
    q_{\Cell}}_{\text{veto}}
    \underbrace{\prod_{\Cell \not \in \VetoCells} \glb 1 - q_{\Cell}
    \grb}_{\text{no veto}}.
    \label{equ:subset_probability}
\end{equation}
Given $\VetoCells$, the loop over far-away cells becomes superfluous,
and the  cell-veto algorithm
can be patched as in \progn{cell-veto(patch)}.

\begin{algorithm}
    \newcommand{\algo}{cell-veto(patch)}
    \captionsetup{margin=0pt,justification=raggedright}
    \begin{center}
        $\begin{array}{ll}
            & \PROCEDURE{\algo}\\
            & \dots\\
            & \COMMENT{treating neighbor and surplus particles as in 
            \progn{cell-veto}}\\
            & \dots\\
            & \IS{\VetoCells}{\text{sampled from $\pi(\VetoCells)$}}\
\COMMENT{see \Eq{equ:subset_probability}}\\
            & \FOR{\Cell \in \VetoCells, \text{non-empty}}\
             \COMMENT{patch of cell-veto loop}\\
            &        \BRACE{
                    \IS{\Upsilon}{\ranb{0, 1}}\\
                    \IS{\xvec''}{\text{(unique) particle $\in \Cell$}}\\
                    \IF{\Upsilon < \fracd{1 -
\expc{-\beta \glb U_{\xvec'' \xvec'}  - U_{\xvec'' \xvec}
\grb }
                                    }{ q_\Cell
} } \GOTO 1 \\
                    }\\ 
            &\IS{\conf}{\SET{ \xvec'} \cup \conf \setminus \SET{\xvec}} \\
           1 & \OUTPUT{\conf}\\
            & \ENDPROCEDURE\
        \end{array}$
    \end{center}
    \caption{\sub{\algo}. Patch of \progn{cell-veto}. The iteration over
    the the far-away cells $\FarCells$ is replaced by the
iteration over the vetoing cells $\VetoCells \subset \FarCells$.}
\label{alg:\algo}
\end{algorithm}

\subsection{Infinitesimal moves, event-chain Monte Carlo}
\label{subsec:event-chain}
In the non-reversible event-chain Monte Carlo
algorithm~\cite{Bernard2009,Michel2014JCP,Krauth2021eventchain}, the move
\begin{equation}
\xvec \to \xvec' = \xvec + \vvec \diff t
\label{equ:InfinitisimalMove}
\end{equation}
is infinitesimal and it is repeated until vetoed by a factor $M$ (in our case a
pair). The particle originally at $\xvec$ thus moves in continuous time along a
ray indicated by the velocity $\vvec$ in \Eq{equ:InfinitisimalMove}. This
velocity can often be taken of unit norm $|\vvec| = 1$ and the inverse move need
not be proposed. After the veto, the other element in the pair factor becomes
the active particle, and this  situation that can be easily generalized
to larger factors describing for example many-body interactions or
groups of atoms indide a pair of molecules
~\cite{Harland2017,Klement2019,Michel2020,Hoellmer2023molecular}.

\begin{figure*}[tb]
	\centering
	\includegraphics{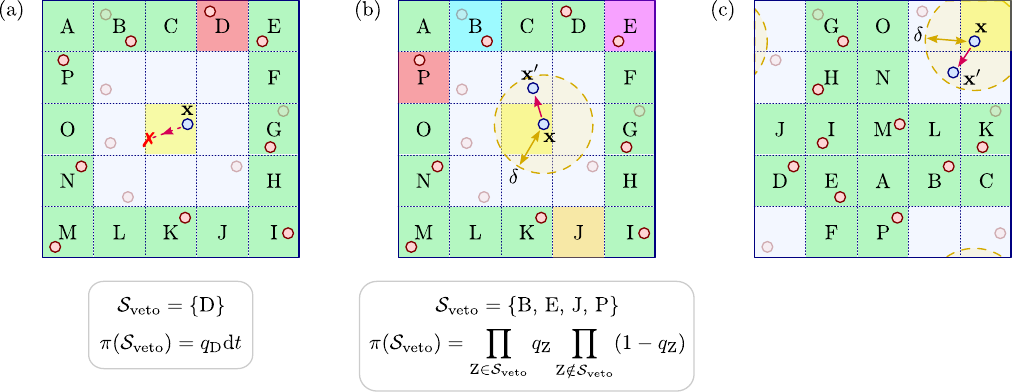}
	\caption{Subsets of cells. \subcap{a} An infinitesimal move, vetoed by at
most a single cell. \subcap{b} A finite move $\xvec \to \xvec'$, vetoed by a
non-trivial subset of cells. \subcap{c} With periodic boundary conditions, the
set $\FarCells$ is defined relative to the active cell.}
	\label{fig:PowersetCells}
\end{figure*}

Within the active cell \quot{0}, a large number of subsequent infinitesimal
moves along the ray are not vetoed by any cell, with the probability
\begin{equation}
1 - \underbrace{\sum_{\Cell \in \FarCells} q_\Cell}_{q_\FarCells},
\end{equation}
giving rise to a waiting-time distribution
\begin{equation}
 \prob(t) \diff t =  q_\FarCells \expb{-q_\FarCells t} \diff t.
\end{equation}
This waiting time for the next cell event can be sampled as $t =
\logc{\ranb{0,1}} / q_\FarCells$. At the event time, the
cell responsible for the veto can be sampled in \bigOb{1} using
Walker's algorithm, as is illustrated in
\figtwo{fig:PowersetCells}{fig:WalkerMethod} for
far-away cells  $\FarCells = \SET{\Atext \TO \Ptext}$.
\begin{figure*}[htb]
    \centering
    \includegraphics{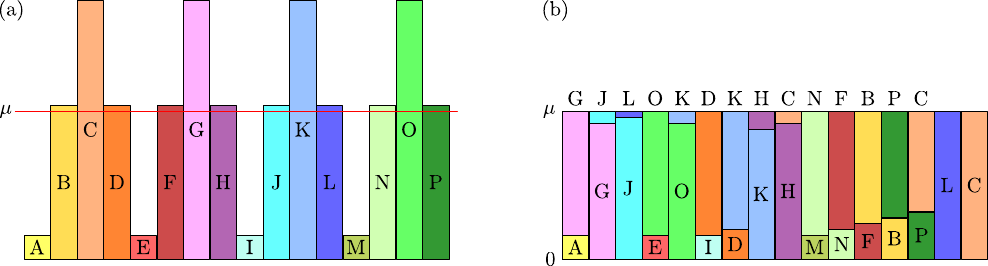}
    \caption{Walker's algorithm for sampling the subset $\VetoCells \subset
\FarCells = \SET{\text{A}, \text{B} \TO \text{P}}$ of vetoing cells
(see \fig{fig:PowersetCells}). The probabilities shown correspond to
$\SET{q_{\text{A}} \TO q_{\text{P}}}$ for infinitesimal moves
or $\SET{\lambda_{\text{A}} \TO \lambda_{\text{P}}}$, for finite moves.
\subcap{a} The probabilities reflect the spatial symmetry of
the cells. \subcap{b} The probabilities are rearranged into a rectangle, with at
most two elements stacked on top of each other. After a \bigOb{|\FarCells|}
preparation (from (a) to (b)), each sample takes two random numbers, so is
\bigOb{1} for $|\FarCells| \to \infty$.
}
\label{fig:WalkerMethod}
\end{figure*}
In a periodic simulation box, the Walker table concerns relative cell
locations, and it is prepared at the beginning of the simulation and then
translates together with the active
particle (see \subfig{fig:PowersetCells}{c}). The method outlined above can  be
integrated into \progn{cell-veto(patch)} for a native constant-time event-chain
Monte Carlo algorithm for the Lennard-Jones model and many other models (see
\fig{fig:ExpData} for run-time data, and  \app{app:Software} for a Python
implementation).

\subsection{Finite displacements}
\label{subsec:powerset-poisson}
For a finite proposed move $\xvec \to \xvec'$, in discrete time and with
finite cell-veto probabilities $q_\Cell$ for $ \Cell \in \FarCells$, more than
one cell can veto, leading to a non-trivial subset $\VetoCells$. While the
cell-accepts are definite, the vetoes are not. They must be confirmed
by a second \quot{pebble}. The set $\VetoCells$, with the probability
distribution given in \Eq{equ:subset_probability},  can be sampled naively by
iterating over the elements of $\FarCells$.
\begin{algorithm}
    \newcommand{\algo}{poisson-veto}
    \captionsetup{margin=0pt,justification=raggedright}
    \begin{center}
        $\begin{array}{ll}
            & \PROCEDURE{\algo}\\
            & \INPUT{\SET{q_\Atext \TO q_\Ptext}}\
             \COMMENT{$q_\Cell$: veto probability of cell $\Cell$}\\
            & \IS{\VetoCells}{\emptyset}\\
            & \FOR{\Cell \in  \SET{\Atext  \TO \Ptext}} \\
            & \BRACE{
             \IS{\lambda_\Cell}{-\logb{1- q_\Cell} }\\
             \IS{t}{ - \logc{\ranb{0,1}} / \lambda_\Cell}\\
             \IF{t < 1}\ \IS{\VetoCells}{\VetoCells \cup \SET{\Cell}} \\
            } \\
            & \OUTPUT{\VetoCells} \\
            & \ENDPROCEDURE\
        \end{array}$
    \end{center}
    \caption{\sub{\algo}. Sampling the set $\VetoCells$ through
$|\FarCells|$ Poisson processes. See patch in
\progn{poisson-veto(patch)} for an equivalent, faster, version
with a single Poisson process.}
\label{alg:\algo}
\end{algorithm}

To sample the set $\VetoCells$ more efficiently, we consider a Poisson process
of intensity $\lambda_\Cell$ in the time interval $[0,1]$, with a distribution
of the number of events as
\begin{equation}
 \prob(\text{$n$ events}) = \fracb{\lambda_\Cell^n}{n!} \expa{-\lambda_\Cell}.
\end{equation}
If we identify $q_\Cell$ with the probability that one or more events take
place in the time interval $[0,1]$, we have:
\begin{align}
 q_\Cell &= \prob(\ge \text{$1$ events}), \\
         & = 1 - \prob(\text{$0$ events}) = 1 - \expa{-\lambda_\Cell},
\end{align}
which gives
$\lambda_\Cell = - \logb{1-q_\Cell}$. The waiting-time distribution of this
Poisson process is
\begin{equation}
 \prob(t) \diff t = \lambda_\Cell \expb{-\lambda_\Cell t} \diff t,
\end{equation}
and events from this process can be sampled as $t =
\logc{\ranb{0,1}}/\lambda_\Cell$. Naively, we might
loop over the cells in $\FCAL$, translate the cell-veto probabilities
$q_\Cell$ into the intensities $\lambda_\Cell$ of associated Poisson processes
and check whether the waiting times are smaller than $1$ (see
\prog{poisson-veto}). Integrated into \prog{cell-veto(patch)}, this exactly
reproduces the factorized Metropolis algorithm.
\begin{figure}[htb]
	\centering
	\includegraphics{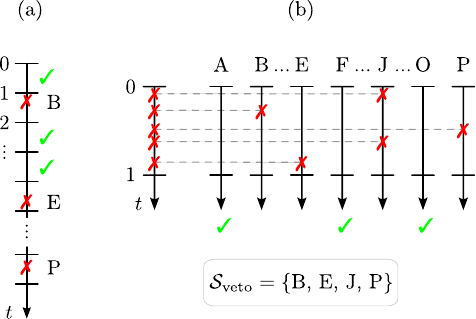}
	\caption{Sampling the veto cells  $\VetoCells$ (see
		\subfig{fig:PowersetCells}{b}).
		\subcap{a} Using subsequent Poisson processes for
		$ \SET{\text{A} \TO\text{P}}$ (see \prog{poisson-veto}.
		\subcap{b} Using a single Poisson
		process, together with Walker's algorithm. Each
		cell may veto repeatedly (see \prog{poisson-veto(patch)}).}
	\label{fig:PowersetPoisson}
\end{figure}

\begin{figure}[htb]
    \centering
    \includegraphics{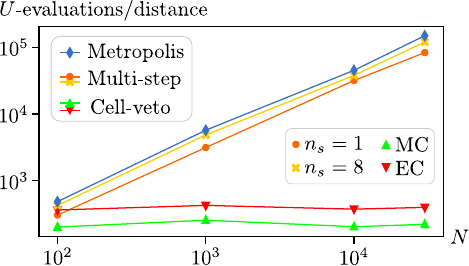}
    \caption{Number of pair-potential evaluations per
unit-distance traveled by a single particle. The results were obtained
using the Python programs of \app{app:Software}. The lines are guides to the
eye. }
    \label{fig:ExpData}
\end{figure}

As the Poisson processes in \prog{poisson-veto} are independent, one may run
them in parallel (see \fig{fig:PowersetPoisson}). A single Poisson process with
intensity $\lambda = \sum_{\Cell \in \FarCells} \lambda_\Cell$ then controls
the time at which one of the $|\FarCells|$ Poisson processes, that is, one of
the cells, holds a veto. Walker's algorithm can again be used to identify this
cell (see \prog{poisson-veto(patch)}). This algorithm avoids the
iteration over the set $\FarCells$ of far-away cells, and as Walker's algorithm
is of complexity \bigOb{1}, the complexity of our algorithm is essentially
\bigOb{|\VetoCells|}. The time interval of the Poisson process may be adjusted
so that $\VetoCells$ contains only few elements. \progg{poisson-veto(patch)}
can be integrated into \progn{cell-veto(patch)} for a native constant-time
Monte Carlo algorithm for the Lennard-Jones model (see \fig{fig:ExpData} for
run-time data, and  \app{app:Software} for a Python implementation).

\begin{algorithm}
    \newcommand{\algo}{poisson-veto(patch)}
    \captionsetup{margin=0pt,justification=raggedright}
    \begin{center}
        $\begin{array}{ll}
            & \PROCEDURE{\algo}\\
            & \INPUT{\SET{q_\Atext \TO q_\Ptext}}\
             \COMMENT{$q_\Cell$: veto probability of cell $\Cell$}\\
            & \IS{\SET{\lambda_\Atext \TO \lambda_\Ptext}}{\SET{-\logb{1-
q_\Atext} \TO -\logb{1 - q_\Ptext}}} \\
            & \IS{\lambda}{\sum_{\Cell \in \FarCells} \lambda_\Cell}\
\COMMENT{total intensity of all Poisson processes}\\
            & \IS{\VetoCells}{\emptyset}\\
            & \IS{t}{0}\\
            & \WHILE{\TRUE}\\
            & \BRACE{
                \IS{t}{t - \logc{\ranb{0,1}} / \lambda}\ \COMMENT{waiting
time to next event}\\
                \IF{t > 1}{\BREAK}\\
                \IS{\Cell}{\sub{walker}(\SET{\lambda_\Atext \TO
\lambda_\Ptext})}\\
                \IS{\VetoCells}{\VetoCells \cup \SET{\Cell}}
                } \\
            & \OUTPUT{\VetoCells}\\
            & \ENDPROCEDURE\
        \end{array}$
    \end{center}
    \caption{\sub{\algo}. Efficient sampling of the set $\VetoCells$
    of veto cells through a single Poisson processes. }
\label{alg:\algo}
\end{algorithm}

\section{Conclusions}

In this paper, we have discussed the role of long-range potentials within
molecular simulation and, especially, within the Monte Carlo approach to
sampling. In a situation where enormous effort has been dedicated to the
efficient evaluation of the total energy and its gradient, we point out that
Monte Carlo sampling of the Boltzmann distribution $\expb{-\beta U}$ does not
generally require knowledge of the total energy $U$. The cell-veto algorithms
that we presented here use a \quot{two-pebble} decision that is akin to the
thinning of Poisson processes, and they can be extended in many directions. In
related research~\cite{Faulkner2018,Hoellmer2020,Hoellmer2023molecular}, we have
shown that the SPC/Fw water model frequently used in biomolecular simulations
can be sampled exactly in the canonical ensemble, without relying on
thermostats, discretization, cutoffs, and grids. It remains to be seen whether
our native methods can be applied more generally to molecular applications and
whether it can solve the problems posed by trucating, smoothing out, or
discretizing  interactions in long-time simulations.
In our examples, the long-range nature of the potentials relate to
the physics of the photon (rather than the meson) for the Coulomb interaction,
and the London dispersion force for the Lennard-Jones
potential of great importance for the physics of interfaces and
surfaces in soft condensed matter.

\acknowledgements
This research was supported by a grant from the Simons Foundation (Grant 839534,
MET). We thank N. Bou-Rabee, T. Schlick, Y. Sugita, and  S. Vionnet for helpful
discussions.

\appendix

\section{Computer programs}
\label{app:Software}
The present work is accompanied by the \texttt{MCLongRange} software package,
which is published as an open-source project under the GNU GPLv3 license.
\texttt{MCLongRange} is available on GitHub as a part of the JeLLyFysh
organization. The package contains Python implementations of many algorithms
discussed in this paper and has been used to produce the numerical results of
\fig{fig:ExpData}. The url of the repository is
\url{https://github.com/jellyfysh/MCLongRange.git}.


\begin{thebibliography}{10}

\bibitem{Schlick2002}
T.~Schlick, {\em {Molecular Modeling and Simulation: An Interdisciplinary
  Guide}}.
\newblock Springer, 2002.

\bibitem{EissfellerOpper1992}
H.~Eissfeller and M.~Opper, ``{New method for studying the dynamics of
  disordered spin systems without finite-size effects},'' {\em Phys. Rev.
  Lett.}, vol.~68, no.~13, p.~2094–2097, 1992.

\bibitem{Jordan2008}
J.~Jordan, R.~Orús, G.~Vidal, F.~Verstraete, and J.~I. Cirac, ``{Classical
  Simulation of Infinite-Size Quantum Lattice Systems in Two Spatial
  Dimensions},'' {\em Phys. Rev. Lett.}, vol.~101, no.~25, 2008.

\bibitem{VanHoucke1}
K.~{Van Houcke}, E.~Kozik, N.~Prokof'ev, and B.~Svistunov {\em Phys. Procedia},
  vol.~6, p.~95, 2010.

\bibitem{VanHouckeEOS}
K.~{Van Houcke}, F.~Werner, E.~Kozik, N.~Prokof'ev, B.~Svistunov, M.~J. Ku,
  A.~S. and  L.W.~Cheuk, A.~Schirotzek, and M.~W. Zwierlein, ``{Feynman
  diagrams versus Fermi-gas Feynman emulator},'' {\em Nature Phys.}, vol.~8,
  p.~366, 2012.

\bibitem{RossiCDet}
R.~Rossi, ``{Determinant Diagrammatic Monte Carlo in the Thermodynamic
  Limit},'' {\em Phys. Rev. Lett.}, vol.~119, p.~045701, 2017.

\bibitem{Levitt2014}
M.~Levitt, ``{Birth and Future of Multiscale Modeling for Macromolecular
  Systems (Nobel Lecture)},'' {\em Angew. Chem. Int. Ed.}, vol.~53,
  p.~10006–10018, Aug. 2014.

\bibitem{Cardy1996}
J.~Cardy, {\em {Scaling and Renormalization in Statistical Physics}}.
\newblock Cambridge University Press, 1996.

\bibitem{AldousDiaconis1986}
D.~Aldous and P.~Diaconis, ``{Shuffling Cards and Stopping Times},'' {\em Am.
  Math. Mon.}, vol.~93, no.~5, pp.~333--348, 1986.

\bibitem{ProppWilson1996}
J.~G. Propp and D.~B. Wilson, ``{Exact sampling with coupled Markov chains and
  applications to statistical mechanics},'' {\em Random Structures Algorithms},
  vol.~9, no.~1-2, pp.~223--252, 1996.

\bibitem{Stone2013}
A.~J. Stone, {\em The Theory of Intermolecular Forces, 2nd edition}.
\newblock Oxford University Press, 2013.

\bibitem{ShawAnton2007}
D.~E. Shaw, M.~M. Deneroff, R.~O. Dror, J.~S. Kuskin, R.~H. Larson, J.~K.
  Salmon, C.~Young, B.~Batson, K.~J. Bowers, J.~C. Chao, M.~P. Eastwood,
  J.~Gagliardo, J.~P. Grossman, C.~R. Ho, D.~J. Ierardi, I.~Kolossv\'{a}ry,
  J.~L. Klepeis, T.~Layman, C.~McLeavey, M.~A. Moraes, R.~Mueller, E.~C.
  Priest, Y.~Shan, J.~Spengler, M.~Theobald, B.~Towles, and S.~C. Wang,
  ``{Anton, a Special-purpose Machine for Molecular Dynamics Simulation},'' in
  {\em Proceedings of the 34th Annual International Symposium on Computer
  Architecture}, ISCA '07, (New York, NY, USA), pp.~1--12, ACM, 2007.

\bibitem{Shaw2010}
D.~E. Shaw, P.~Maragakis, K.~Lindorff-Larsen, S.~Piana, R.~O. Dror, M.~P.
  Eastwood, J.~A. Bank, J.~M. Jumper, J.~K. Salmon, Y.~Shan, and W.~Wriggers,
  ``{Atomic-Level Characterization of the Structural Dynamics of Proteins},''
  {\em Science}, vol.~330, no.~6002, pp.~341--346, 2010.

\bibitem{greengard1987}
L.~Greengard and V.~Rokhlin, ``{A fast algorithm for particle simulations},''
  {\em J. Comput. Phys.}, vol.~73, no.~2, pp.~325--348, 1987.

\bibitem{Smit1991}
B.~Smit and D.~Frenkel, ``{Vapor{\textendash}liquid equilibria of the
  two-dimensional Lennard-Jones fluid(s)},'' {\em J. Chem. Phys.}, vol.~94,
  no.~8, pp.~5663--5668, 1991.

\bibitem{Smit1992}
B.~Smit, ``{Phase diagrams of Lennard-Jones fluids},'' {\em J. Chem. Phys.},
  vol.~96, no.~11, pp.~8639--8640, 1992.

\bibitem{LeiBykov2005}
Y.~A. Lei, T.~Bykov, S.~Yoo, and X.~C. Zeng, ``{The Tolman Length: Is It
  Positive or Negative?},'' {\em J. Am. Chem. Soc.}, vol.~127, no.~44,
  p.~15346–15347, 2005.

\bibitem{Troester2012}
A.~Tr\"{o}ster, M.~Oettel, B.~Block, P.~Virnau, and K.~Binder, ``{Numerical
  approaches to determine the interface tension of curved interfaces from free
  energy calculations},'' {\em J. Chem. Phys.}, vol.~136, no.~6, 2012.

\bibitem{Tempra2022}
C.~Tempra, O.~H.~S. Ollila, and M.~Javanainen, ``{Accurate Simulations of Lipid
  Monolayers Require a Water Model with Correct Surface Tension},'' {\em J.
  Chem. Theory Comput.}, vol.~18, no.~3, p.~1862–1869, 2022.

\bibitem{Yu2021}
Y.~Yu, A.~Kr\"{a}mer, R.~M. Venable, B.~R. Brooks, J.~B. Klauda, and R.~W.
  Pastor, ``{CHARMM36 Lipid Force Field with Explicit Treatment of Long-Range
  Dispersion: Parametrization and Validation for Phosphatidylethanolamine,
  Phosphatidylglycerol, and Ether Lipids},'' {\em J. Chem. Theory Comput.},
  vol.~17, no.~3, p.~1581–1595, 2021.

\bibitem{SMAC}
W.~Krauth, {\em {Statistical Mechanics: Algorithms and Computations}}.
\newblock Oxford University Press, 2006.

\bibitem{KapferKrauth2016}
S.~C. Kapfer and W.~Krauth, ``{Cell-veto Monte Carlo algorithm for long-range
  systems},'' {\em Phys. Rev. E}, vol.~94, p.~031302, 2016.

\bibitem{Michel2014JCP}
M.~{Michel}, S.~C. {Kapfer}, and W.~{Krauth}, ``{Generalized event-chain Monte
  Carlo: Constructing rejection-free global-balance algorithms from
  infinitesimal steps},'' {\em J. Chem. Phys.}, vol.~140, no.~5, p.~054116,
  2014.

\bibitem{Tartero2024}
G.~Tartero and W.~Krauth, ``{Concepts in Monte Carlo sampling},'' {\em Am. J.
  Phys.}, vol.~92, no.~1, pp.~65--77, 2024.

\bibitem{LewisShedler1979}
P.~A.~W. Lewis and G.~S. Shedler, ``{Simulation of nonhomogeneous Poisson
  processes by thinning},'' {\em Nav. Res. Logist. Q.}, vol.~26, no.~3,
  pp.~403--413, 1979.

\bibitem{MuellerJanke2023}
F.~M\"{u}ller, H.~Christiansen, S.~Schnabel, and W.~Janke, ``{Fast,
  Hierarchical, and Adaptive Algorithm for Metropolis Monte Carlo Simulations
  of Long-Range Interacting Systems},'' {\em Phys. Rev. X.}, vol.~13, no.~3,
  2023.

\bibitem{Berne2002}
B.~Hetényi, K.~Bernacki, and B.~J. Berne, ``{Multiple “time step” Monte
  Carlo},'' {\em J. Chem. Phys.}, vol.~117, no.~18, pp.~8203--8207, 2002.

\bibitem{MichelTanDeng2019}
M.~Michel, X.~Tan, and Y.~Deng, ``{Clock Monte Carlo methods},'' {\em Phys.
  Rev. E}, vol.~99, p.~010105, 2019.

\bibitem{Hoellmer2023molecular}
P.~H\"{o}llmer, A.~C. Maggs, and W.~Krauth, ``{Molecular simulation from modern
  statistics: Continuous-time, continuous-space, exact}.''
  \url{https://arxiv.org/abs/2305.02979}, 2023.

\bibitem{Harland2017}
J.~Harland, M.~Michel, T.~A. Kampmann, and J.~Kierfeld, ``{Event-chain Monte
  Carlo algorithms for three- and many-particle interactions},'' {\em EPL},
  vol.~117, no.~3, p.~30001, 2017.

\bibitem{LennardJones1931}
J.~E. Lennard-Jones, ``{Cohesion},'' {\em Proc. Phys. Soc.}, vol.~43, no.~5,
  p.~461, 1931.

\bibitem{Faulkner2018}
M.~F. Faulkner, L.~Qin, A.~C. Maggs, and W.~Krauth, ``{All-atom computations
  with irreversible Markov chains},'' {\em J. Chem. Phys.}, vol.~149, no.~6,
  p.~064113, 2018.

\bibitem{DuKePeRo1987}
S.~Duane, A.~D. Kennedy, B.~J. Pendleton, and D.~Roweth, ``{Hybrid
  Monte-Carlo},'' {\em Phys Lett B}, vol.~195, pp.~216--222, 1987.

\bibitem{Neal2011}
R.~M. Neal, ``{MCMC using Hamiltonian dynamics},'' in {\em {Handbook of Markov
  Chain Monte Carlo}} (S.~Brooks, A.~Gelman, G.~Jones, and X.-L. Meng, eds.),
  pp.~113--162, Chapman and Hall/CRC, 2011.

\bibitem{BouRabee2018}
N.~Bou-Rabee and J.~M. Sanz-Serna, ``{Geometric integrators and the Hamiltonian
  Monte Carlo Method},'' {\em Acta Numer.}, vol.~27, pp.~113--206, 2018.

\bibitem{Krauth2021eventchain}
W.~Krauth, ``{Event-Chain Monte Carlo: Foundations, Applications, and
  Prospects},'' {\em Front. Phys.}, vol.~9, p.~229, 2021.

\bibitem{Walker1977AnEfficientMethod}
A.~J. Walker, ``{An Efficient Method for Generating Discrete Random Variables
  with General Distributions},'' {\em ACM Trans. Math. Softw.}, vol.~3, no.~3,
  pp.~253--256, 1977.

\bibitem{Bernard2009}
E.~P. Bernard, W.~Krauth, and D.~B. Wilson, ``{Event-chain Monte Carlo
  algorithms for hard-sphere systems},'' {\em Phys. Rev. E}, vol.~80,
  p.~056704, 2009.

\bibitem{Klement2019}
M.~Klement and M.~Engel, ``{Efficient equilibration of hard spheres with
  Newtonian event chains},'' {\em J. Chem. Phys.}, vol.~150, no.~17, p.~174108,
  2019.

\bibitem{Michel2020}
M.~Michel, A.~Durmus, and S.~S{\'{e}}n{\'{e}}cal, ``{Forward Event-Chain Monte
  Carlo: Fast Sampling by Randomness Control in Irreversible Markov Chains},''
  {\em J. Comput. Graph. Stat.}, vol.~29, no.~4, pp.~689--702, 2020.

\bibitem{Hoellmer2020}
P.~H\"{o}llmer, L.~Qin, M.~F. Faulkner, A.~C. Maggs, and W.~Krauth,
  ``{JeLLyFysh-Version1.0~{\textemdash} a Python application for all-atom
  event-chain Monte Carlo},'' {\em Comput. Phys. Commun.}, vol.~253, p.~107168,
  2020.

\end{thebibliography}
\end{document}